\documentclass[12pt]{article}
\usepackage{graphics,psfrag}
\usepackage{amsthm,amssymb,epsfig,amsmath,cite}
\usepackage{euscript,array}

\newcounter{multieqs}

\newcommand{\be}{\begin{equation}}
\newcommand{\ee}{\end{equation}}
\newcommand{\eq}[1]{(\ref{#1})}

\newcommand{\bm}[1]{\mbox{\boldmath $#1$}}
\newcommand{\rf}[1]{(\ref{#1})}

\def\bd{\begin{document}}
\def\ed{\end{document}}
\def\nn{\nonumber}
\def\bea{\begin{eqnarray}}
\def\eea{\end{eqnarray}}
\let\bm=\bibitem
\let\la=\label

\def\npb#1#2#3{Nucl. Phys. {\bf{B#1}} #3 (#2)}
\def\plb#1#2#3{Phys. Lett. {\bf{#1B}} #3 (#2)}
\def\prl#1#2#3{Phys. Rev. Lett. {\bf{#1}} #3 (#2)}
\def\prd#1#2#3{Phys. Rev. {D \bf{#1}} #3 (#2)}
\def\cmp#1#2#3{Comm. Math. Phys. {\bf{#1}} #3 (#2)}
\def\cqg#1#2#3{Class. Quantum Grav. {\bf{#1}} #3 (#2)}
\def\nppsa#1#2#3{Nucl. Phys. B (Proc. Suppl.) {\bf{#1A}}#3 (#2)}
\def\ap#1#2#3{Ann. of Phys. {\bf{#1}} #3 (#2)}
\def\ijmp#1#2#3{Int. J. Mod. Phys. {\bf{A#1}} #3 (#2)}
\def\rmp#1#2#3{Rev. Mod. Phys. {\bf{#1}} #3 (#2)}
\def\mpla#1#2#3{Mod. Phys. Lett. {\bf A#1} #3 (#2)}
\def\jhep#1#2#3{J. High Energy Phys. {\bf #1} #3 (#2)}
\def\atmp#1#2#3{Adv. Theor. Math. Phys. {\bf #1} #3 (#2)}

\newcommand{\EQ}[1]{\begin{equation} #1 \end{equation}}
\newcommand{\AL}[1]{\begin{subequations}\begin{align} #1 \end{align}
\end{subequations}}
\newcommand{\SP}[1]{\begin{equation}\begin{split} #1 \end{split}\end{equation}}
\newcommand{\ALAT}[2]{\begin{subequations}\begin{alignat}{#1} #2 
\end{alignat}\end{subequations}}
\def\beqa{\begin{eqnarray}} 
\def\eeqa{\end{eqnarray}} 
\def\beq{\begin{equation}} 
\def\eeq{\end{equation}} 

\def\N{{\cal N}}
\def\sst{\scriptscriptstyle}
\def\thetabar{\bar\theta}
\def\Tr{{\rm Tr}}
\def\one{\mbox{1 \kern-.59em {\rm l}}}

\def\a{\alpha}      \def\da{{\dot\alpha}}  
\def\b{\beta}       \def\db{{\dot\beta}}  
\def\c{\gamma}  \def\C{\Gamma}  \def\dc{{\dot\gamma}}  
\def\d{\delta}  \def\D{\Delta}  \def\ddt{\dot\delta}  
\def\e{\epsilon}        \def\vare{\varepsilon}  
\def\f{\phi}    \def\F{\Phi}    \def\vvf{\f}  
\def\h{\eta}  
\def\k{\kappa}  
\def\l{\lambda} \def\L{\Lambda}  
\def\m{\mu} \def\n{\nu}  
\def\o{\omega}  
\def\p{\pi} \def\P{\Pi}  
\def\r{\rho}  
\def\s{\sigma}  \def\S{\Sigma}  
\def\t{\tau}  
\def\th{\theta} \def\Th{\Theta} \def\vth{\vartheta}  
\def\X{\Xeta}  
\def\z{\zeta}  
  
\def\cA{{\cal A}} \def\cB{{\cal B}} \def\cC{{\cal C}}  
\def\cD{{\cal D}} \def\cE{{\cal E}} \def\cF{{\cal F}}  
\def\cG{{\cal G}} \def\cH{{\cal H}} \def\cI{{\cal I}}  
\def\cJ{{\cal J}} \def\cK{{\cal K}} \def\cL{{\cal L}}  
\def\cM{{\cal M}} \def\cN{{\cal N}} \def\cO{{\cal O}}  
\def\cP{{\cal P}} \def\cQ{{\cal Q}} \def\cR{{\cal R}}  
\def\cS{{\cal S}} \def\cT{{\cal T}} \def\cU{{\cal U}}  
\def\cV{{\cal V}} \def\cW{{\cal W}} \def\cX{{\cal X}}  
\def\cY{{\cal Y}} \def\cZ{{\cal Z}}

\def\ua{\underline{\alpha}}  
\def\ub{\underline{\phantom{\alpha}}\!\!\!\beta}  
\def\uc{\underline{\phantom{\alpha}}\!\!\!\gamma}  
\def\um{\underline{\mu}}  
\def\ud{\underline\delta}  
\def\ue{\underline\epsilon}  
\def\una{\underline a}\def\unA{\underline A}  
\def\unb{\underline b}\def\unB{\underline B}  
\def\unc{\underline c}\def\unC{\underline C}  
\def\und{\underline d}\def\unD{\underline D}  
\def\une{\underline e}\def\unE{\underline E}  
\def\unf{\underline{\phantom{e}}\!\!\!\! f}\def\unF{\underline F}  
\def\unm{\underline m}\def\unM{\underline M}  
\def\unn{\underline n}\def\unN{\underline N}  
\def\unp{\underline{\phantom{a}}\!\!\! p}\def\unP{\underline P}  
\def\unq{\underline{\phantom{a}}\!\!\! q}  
\def\unQ{\underline{\phantom{A}}\!\!\!\! Q}  
\def\unH{\underline{H}}  
  
\def\As {{A \hspace{-6.4pt} \slash}\;}  
\def\bs {{b \hspace{-6.4pt} \slash}\;}  
\def\Ds {{D \hspace{-6.4pt} \slash}\;}  
\def\ds {{\del \hspace{-6.4pt} \slash}\;}  
\def\ks {{ k \hspace{-6.4pt} \slash}\;}  
\def\ps {{p \hspace{-6.4pt} \slash}\;}  
\def\pas {{{p_1} \hspace{-6.4pt} \slash}\;}  
\def\pbs {{{p_2} \hspace{-6.4pt} \slash}\;}  
  
\def\Ah{\hat{A}}  
\def\Dh{\hat{D}}  
\def\Fh{\hat{F}}  
\def\Vh{\hat{V}}  
\def\Xh{\hat{X}}  
\def\ah{\hat{a}}  
\def\xh{\hat{x}}  
\def\yh{\hat{y}}  
\def\ph{\hat{p}}  
\def\xih{\hat{\xi}} 
\def\phih{\hat{\phi}} 
\def\psih{\hat{\psi}}
\def\Deltah{\hat{\Delta}}
  
\def\psit{\tilde{\psi}}  
\def\Psit{\tilde{\Psi}}   
\def\Psibt{\tilde{\bar{Psi}}}  

\def\Phit{\tilde{\Phi}}   
\def\Phitb{\overline{\tilde{Phi}}}  
\def\tht{\tilde{\th}}  
   
\def\At{\tilde{A}}  
\def\Qt{\tilde{Q}}  
\def\Rt{\tilde{R}}  
\def\Nt{\tilde{N}}  
 
\def\at{\tilde{a}}  
\def\st{\tilde{s}}  
\def\ft{\tilde{f}}  
\def\pt{\tilde{p}}  
\def\qt{\tilde{q}}  
\def\vt{\tilde{v}}  
\def\nt{\tilde{n}}  
  
\def\delb{\overline{\partial}}  
\def\thb{\overline{\theta}}
\def\mub{{\overline \mu}}
\def\lamb{{\overline \l}}
\def\psib{{\overline \psi}}
\def\sb{{\overline \sigma}}
\def\xib{{\overline \xi}}
\def\chib{{\overline \chi}}

\def\Phib{\overline{\Phi}}
\def\Lamb{\overline{\Lambda}}

\def\Ab{{\overline A}} \def\Bb{{\overline B}} \def\Cb{{\overline C}}  
\def\Db{{\overline D}} \def\Eb{{\overline E}} \def\Fb{{\overline F}}  
\def\Gb{{\overline G}} \def\Hb{{\overline H}} \def\Ib{{\overline I}}  
\def\Jb{{\overline J}} \def\Kb{{\overline K}} \def\Lb{{\overline L}}  
\def\Mb{{\overline M}} \def\Nb{{\overline N}} \def\Ob{{\overline O}}  
\def\Pb{{\overline P}} \def\Qb{{\overline Q}} \def\Rb{{\overline R}}  
\def\Sb{{\overline S}} \def\Tb{{\overline T}} \def\Ub{{\overline U}}  
\def\Vb{{\overline V}} \def\Wb{{\overline W}} \def\Xb{{\overline X}}  
\def\Yb{{\overline Y}} \def\Zb{{\overline Z}}  

\def\fb{{\overline f}}
\def\gb{{\overline g}}
\def\mb{{\overline m}}
\def\lb{{\overline l}}
\def\yb{{\overline y}}

\def\ba{{\bf a}} 
\def\bk{{\bf k}}  
\def\bl{{\bf l}}  
\def\bp{{\bf p}}  
\def\bq{{\bf q}}  
\def\br{{\bf r}}
\def\bt{{\bf t}}
\def\bu{{\bf u}}
\def\bv{{\bf v}}
\def\bx{{\bf x}}  
\def\by{{\bf y}}  
\def\bR{{\bf R}}  
\def\bV{{\bf V}}

\def\bone{{\bf 1}}  

\def\va{{\vec a}}
\def\vp{{\vec p}}
\def\vq{{\vec q}}
\def\vx{{\vec x}}
\def\vu{{\vec u}}
\def\vv{{\vec v}}

\def\vs{{\vec \sigma}}
\def\vtau{{\vec \tau}}

\newcommand{\ov}[1]{\overrightarrow{#1}}
  
\def\d{\delta}\def\D{\Delta}\def\ddt{\dot\delta}  
  
\def\pa{\partial} \def\del{\partial}  
\def\xx{\times}  
\def\uno{\mbox{1 \kern-.59em {\rm l}}}    
  
\def\trp{^{\top}}  
\def\inv{^{-1}}  
\def\dag{{^{\dagger}}}  
\def\pr{^{\prime}}  
  
\def\rar{\rightarrow}  
\def\lar{\leftarrow}  
\def\lrar{\leftrightarrow}  
  
\newcommand{\0}{\,\!}      
\def\one{1\!\!1\,\,}  
\newcommand{\im}{\,\mathrm{i}\,}  
\newcommand{\ep}{\,\mathrm{e}\,}
\def\jm{\jmath}  
  
\newcommand{\tr}{\mbox{tr}}  
\newcommand{\slsh}[1]{/ \!\!\!\! #1}  
  
\def\vac{|0\rangle}  
\def\lvac{\langle 0|}  
  
\def\hlf{\frac{1}{2}}  
\def\ove#1{\frac{1}{#1}}  

\def\Box{\square}  
\def\ZZ{\mathbb{Z}}  
\def\RR{\mathbb{R}}
\def\CC{\mathbb{C}}
\def\bb#1{{\bf #1}}  
\def\bcomment#1{}  
\def\bfhat#1{{\bf \hat{#1}}}  
\def\VEV#1{\left\langle #1\right\rangle}  

\newcommand{\ex}[1]{{\rm e}^{#1}} \def\ii{{\rm i}}  

\newcommand{\lrbrk}[1]{\left(#1\right)}
\newcommand{\sfrac}[2]{{\textstyle\frac{#1}{#2}}}
 
\def\stw{{\sqrt{2}}}

\def\rf {{\rm f}}
\def\ri {{\rm i}}
\def\rs {{\scriptscriptstyle \rm S}}
\def\rt {{\scriptscriptstyle \rm T}}

\def\rQ {{\scriptscriptstyle \rm \cQ}}
\def\rR {{\scriptscriptstyle \rm \cR}}

\def\cQb{{\cal \Qb}}
\def\cRb{{\cal \Rb}}
\def\cWb{{\cal \Wb}}

\def\fd {{\rm N}}
\def\afd {{\overline{\rm N}}}

\font\myBB=msbm10 at 18pt
\def\BB#1{\hbox{\myBB#1}}

\newcommand{\sect}[1]{\underline{\it #1}}

\begin{document}

\begin{flushright}
hep-th/0508055\\
ITP--UH--12/05
\end{flushright}

\vspace{20pt}

\begin{center}

{\Large \bf	Emergence of Time from Dimensional Reduction \\[6pt]
		in Noncommutative Geometry} 
\vspace{30pt}

{\large Chong-Sun Chu~$^1$\ \ and \ \ Olaf Lechtenfeld~$^2$} \\[24pt]

{\em ${}^1\ $ 
     Centre for Particle Theory and Department of Mathematics  \\
     University of Durham, Durham, DH1 3LE, United Kingdom}\\
{\tt chong-sun.chu@durham.ac.uk} \\[12pt]     

{\em ${}^2\ $
     Institut f\"ur Theoretische Physik, Universit\"at Hannover  \\
     Appelstra\ss{}e 2, D-30167 Hannover, Germany}\\
{\tt lechtenf@itp.uni-hannover.de}

\vspace{80pt}
{\bf Abstract}

\end{center}
\noindent
By considering a new form of dimensional reduction for noncommutative 
field theory, we show that the signature of spacetime may be changed.
In particular, it is demonstrated that a temporal dimension can emerge 
from a purely Euclidean geometry. We suggest that this mechanism may hint 
at the origin of time in the fundamental theory of quantum gravity.

\setcounter{page}0
\newpage

\section{Introduction}
 
In general, given an action with the kinetic term 
\be
S= \int h\textsc{}^{\m\n}\,\del_\m\psi\,\del_\n \psi 
\ee
for a real scalar field $\psi$,
one can read off the metric $h_{\m\n}$ of the underlying space directly
by taking the inverse of the matrix $(h^{\m\n})$. If Lorentzian, the metric 
yields a light cone, determines the field propagation 
and gives rises to the concept of macrocausality. Performing a canonical
quantization, one may obtain from the vanishing of the commutator of
observables a microcausality condition. 
It is well known that ordinary local interaction in quantum field theory 
cannot modify the lightcone. It is also well known that while conventional 
dimensional reduction can change a Minkowskian spacetime to a Euclidean one 
by a simple reduction on time, to generate a time dimension from a purely
Euclidean space is impossible.

The situation is drastically different for field theory in noncommutative 
space (for reviews see~\cite{review1,review2,review3}).  Recently, it has
been shown that the microcausality  of noncommutative field theory is
generally modified from a lightcone to a lightwedge~\cite{lclw1,lclw2}. 
This modification is due to the highly nonlocal 
nature of the noncommutative interaction.
In this letter we demonstrate another effect of noncommutative geometry 
on the lightcone. We will show that, by 
performing a more general form of dimensional reduction, one can 
generate a time dimension from a purely Euclidean noncommutative space. 

The change of signature in noncommutative geometry hints at a novel
possibility to explain the origin of time.
In our framework, time is not fundamental but a concept emergent from space.
We emphasize that noncommutativity plays a crucial role for our mechanism 
to work. It has been argued by DeWitt that quantum gravity has an uncertainty
principle which prevents one from measuring positions to better
than Planck length accuracy~\cite{dewitt}.
It is natural to expect that noncommutative geometry plays an 
important role in this realm. Therefore, one may speculate
the emergence of time to be an intrinsic  property of quantum gravity. 
Our results suggest a fundamental ``timeless'' formulation of quantum gravity.
Even more, it has been proposed
that both space and time are not fundamental and will have to be replaced 
by something more sophisticated~\cite{cole}. 
It is likely that the fundamental theory of quantum gravity is ``pointless'', 
meaning that both space and time are secondary, derived or effective
notions.\footnote{
A step in this direction is provided by the IKKT matrix model
\cite{aikkt} where it has been argued that a spacetime continuum is 
generated dynamically. However a Minkowski signature is assumed there.}
In this letter, we provide a simple mechanism towards a realization of 
this speculation.

\section{Dimensional reduction in the commutative case}
 
Let us start with a review of the commutative case. 
Consider a commutative gauge theory with the action
\be \label{S3}
S_{D+1}=\int\tr\,\Bigl[(\Dh_\m \phih)^2 + \sfrac{1}{2}\Fh_{\m\n}^2 \Bigr]\quad, 
\ee
where
\be
\Dh_\m \phih = \del_\m \phih +[\Ah_\m, \phih] \quad, \qquad 
\Fh_{\m\n} = \del_\m \Ah_\n -\del_\n \Ah_\m +[\Ah_\m\,,\Ah_\n] \quad,
\ee
and the fields  $\phih$ and $\Ah_\m$ are $N \times N$ real matrices.
The action is GL$(N,\RR)$ invariant under the gauge transformation
\be \label{g-transf}
\phih \rightarrow g^{-1} \phih\, g \qquad \textrm{and} \qquad 
\Ah_\m \rightarrow g^{-1} \Ah_\m g + g^{-1} \del_\m g
\qquad\textrm{for}\quad g\in\textrm{GL}(N,\RR) \quad.
\ee
We will be interested in the Euclidean space with metric  
$g_{\m\n} = \d_{\m\n}$. Let the coordinates be
\be
(x^\m) = (x^0, x^1, \cdots, x^{D-1}, z) = (\{x^i\}, z) =: (\vx,z) \quad,
\ee  
and $z$ is the coordinate we will take to reduce the theory.

Usually the dimensional reduction takes the form that 
all the fields are declared independent of $z$:
\be \label{naive}
\phih = \phi(\vx) \qquad\textrm{and}\qquad \Ah_\m = A_\m(\vx) \quad.
\ee 
This leads to the action
\footnote{We drop the unimportant normalization $\int dz$.}
\be \label{S2}
S_D = \int \tr\,\Bigl[ (D_i \phi)^2 + (D_i \l)^2 +[\phi\,,\l]^2 +
\sfrac{1}{2}F_{ij}^2 \Bigr] \quad,
\ee
where 
\be 
\l:=A_z(\vx)\quad,\qquad
F_{ij}: = \del_i A_j -\del_j A_i +[A_i\,,A_j] \quad,\qquad 
D_i := \del_i +[A_i\,,\cdot\;] \quad. 
\ee 

Let us consider a more general reduction of the ``twisted'' form
\be \label{red-gen}
\phih = U(\vx,z)\,\phi(\vx)\,U^{-1}(\vx,z) \qquad\textrm{and}\qquad 
\Ah_\m = U(\vx,z)\,A_\m(\vx)\,U^{-1}(\vx,z) 
\ee
for $U(\vx,z) \in\textrm{GL}(N,\RR)$.
This is gauge equivalent to 
\be \label{red-A}
\phih = \phi(\vx)\qquad\textrm{and}\qquad 
\Ah_\m = A_\m(\vx) + \del_\m U U^{-1}(\vx,z) \quad,
\ee
which is just a reduction in the presence of a 
vacuum background gauge field 
\be\label{A_B}
\Ah_\m^{(B)} = \del_\m U U^{-1} \quad.
\ee
The choice of $U$ is not arbitrary. To see this, note that
our ansatz \eq{red-gen} gives
\bea
U^{-1} \Fh_{ij} U &=& 
F_{ij} +[U^{-1}\del_i U,A_j] - [U^{-1}\del_j U,A_i]\quad, \label{F1} \\
U^{-1}\Fh_{iz} U &=& 
D_i \l + [ U^{-1}\del_i U, \l]- [U^{-1}\del_z U,A_i]\quad, \\
U^{-1}\Dh_i \phih U &=& D_i \phi +[U^{-1}\del_i U,\phi]\quad, \\
U^{-1} \Dh_z \phih\,U &=& [\l\,,\phi] +[U^{-1}\del_z U,\phi]\quad .\label{F4}
\eea
To be a consistent reduction, the action should be independent of $z$
when these expressions are substituted. 
To achieve this, we want \eq{F1}--\eq{F4} to be independent of~$z$.
Without making any special assumption on the fields $\phi$ and $A_\m$,
a sufficient condition is
\be \label{UU}
U^{-1} \del_\m  U = h_\m(\vx,z) + k_\m(\vx) \qquad 
\textrm{with $h_\m$ being central} \quad.
\ee 
For example, one may take $U =U(\vx)$. In this
case, the background gauge field \eq{A_B} is $z$-independent. 
As a result of \eq{UU}, it is easy to verify that the reduced
action is given by \eq{S2} plus a couple of additional terms that arise
due to the background gauge configuration.

To obtain the propagator, gauge fixing is 
necessary to make sure that the quadratic part of \eq{S3} or \eq{S2}
is invertible. A convenient choice of the gauge-fixing condition is
\be\label{gf3}
\del_\m \Ah^\m =0 \quad. 
\ee
Including the gauge-fixing term $S_{\rm g.f.}= \int \tr(\del_\m \Ah^\m)^2$, 
this leads to the gauge-fixed action,
\be\label{S3'}
S_{D+1}':= S_{D+1}+ S_{\rm g.f.} 
= \int \tr\,\Bigl[ -\phih \Deltah \phih - \delta^{\m\n}\Ah_\m \Deltah \Ah_\n 
+ \cdots \Bigr]
\qquad \textrm{with} \quad  \Deltah := \del_\mu \del^\mu \quad,
\ee
where we wrote out only the kinetic terms. Note that 
\be \label{gf2}
U^{-1} \del_\m \Ah^\m U =
\del_i A^i +[U^{-1} \del_i U, A^i] + [U^{-1}\del_z  U, \l] =
\del_i A^i +[k_i\,,A^i] + [k_z\,,\l] \quad,
\ee
and so the gauge-fixing condition reduces to a lower-dimensional one
provided that \eq{UU} is satisfied.
The reduced gauge-fixed action takes the form
\be
S_D'= \int \tr\,\Bigl[ -\phi \D \phi - A_i \D A_i
- \l \D \l + \cdots \Bigr] \quad ,
\qquad\textrm{where}\quad \D := \del_i \del_i \quad.
\ee 
It is clear that the metric gets reduced as follows,
\be
\d_{\m\n} \rightarrow \d_{ij} \quad,
\ee
and remains Euclidean. 
This conclusion will not be affected by using another choice of gauge fixing.

\section{Noncommutative reduction and generation of time}

We are interested in dimensional reduction in a noncommutative Euclidean
space. Note that one cannot take the fields $\phih$ and $\Ah_\m$ to be
real any more since, in the presence of the star product, this is not
compatible with the gauge transformation \eq{g-transf}. In fact, the
gauge group must now be extended to $\textrm{GL}_\star(N,\CC)$, and the
action $S_{D+1}$ in \eq{S3} ceases to be real. This does not bother us
as the latter only serves to formulate the extremum principle for the
equations of motion. 

Normally one would like to associate the action with quantum amplitudes 
via the phase $e^{\im S}$.
However, a space without time
does not provide an arena for standard quantum mechanics, and one should not 
expect the action to do more than generating the equations of motion. 
After reduction, we 
will arrive at a lower-dimensional world with time. Therefore, 
the reduced action should be real in order to accomodate quantum processes. 
For the original Euclidean theory, one may consider a 
path integral defined by $Z:= \int [D \phih] \, e^{\im {\rm Re} S_{D+1}}$.
Upon dimensional reduction, this partition function give rises to the 
standard path integral for the lower dimensional Minkowskian theory. 
It would be interesting to understand further the physical properties of~$Z$. 

Now back to the dimensional reduction. 
We assume that there is at least one coordinate
which is commuting, say $z$, and we will reduce on this coordinate. For
simplicity let us 
take $N=1$ and consider the case of $D=2$, i.e.
$(x^\m)= (t,x,z)=(\vx,z)$ with Euclidean metric.  The generalization
to the nonabelian case and to higher dimensions is straightforward. 
The noncommutativity is given by
\be
[t,x] = \im \th \qquad\textrm{and $z$ being central}\quad.
\ee

Since GL${}_\star(1,\CC)$ is infinite-dimensional,
the condition \eq{UU} becomes much 
more restrictive. In particular, in order  to be central, 
$h_\m$ must not depend on $t$ or $x$, enforcing~\footnote{
The star product and star exponential are understood below.
}
\be \label{UU-3}
U^{-1}\del_\m U = h_\m(z) + k_\m(\vx) \quad.
\ee 
We will now show that these equations essentially fix the form of $U$ to be 
\be \label{U-gen}
U = U_0 := e^{z f(t,x)} \quad,\qquad \textrm{where $f(t,x) = \a t + \b x$}
\qquad\textrm{and $\a,\b$ are constants} \quad.
\ee
The proof is straightforward. First, from the consistency of the 
$\mu{=}i$ equations, one easily see that $k_i$ has to be a pure gauge, i.e.
\be
k_i =  W^{-1} \del_i W \qquad\textrm{for some $W(\vx)$}\quad.
\ee
Second, the $\mu{=}i$ equations are compatible with the $\mu{=}z$ equation
only if
\be
\del_z h_i = c_i = \del_i k_z + [k_i\,,k_z] 
\qquad\textrm{for constants $c_i$} \quad,
\ee
which implies that
\be
h_i = c_i z + d_i \qquad\textrm{and}\qquad 
k_z = W^{-1} (c_i x^i + d)\,W
\ee
with further constants $d_i$ and $d$. The function $h_z$ is unconstrained.
It follows that the corresponding $U$ factorizes as
\be
U = V(z)\,U_0(z,\vx)\,W(\vx) \qquad\textrm{with}\quad
U_0 = \ep^{c_i x^i z + d_i x^i + d z} \quad,
\ee
so that we have
\be
h_i = U_0^{-1} \del_i U_0 \qquad\textrm{and}\qquad h_z = V^{-1} \del_z
V\quad .
\ee
Clearly, the factors $V$ and $W$ can be absorbed into redefinitions of
$\phi$, $A_i$, $\l$ and $\phih$, $\Ah_i$, respectively. For the same reason
we can drop the constants $d_i$ and~$d$. Therefore, the essential part
of~$U$ is the one that entangles the $z$ and $x^i$ dependence, as claimed
in~\eq{U-gen}.
The corresponding background gauge field is
\be \label{A-bkgd}
\Ah^{(B)}_t =\a z \quad, \qquad \Ah^{(B)}_x =\b z \quad, \qquad
\Ah^{(B)}_z =\a t+\b x \quad,
\ee
and the gauge-fixing condition \eq{gf2} reads
\be
\del_i A^i + [\Ah^{(B)}_z,\l] =0 \quad.
\ee
So far, we have shown that for the noncommutative gauge theory \eq{S3}, 
the twisted reduction defined in~\eq{red-gen} is essentially determined 
by~\eq{U-gen}. It corresponds to having a linear gauge
potential background~\eq{A-bkgd}. Therefore, we may call our reduction
a ``linear background reduction''.

For the $U$ of \eq{U-gen}, we have
\bea
U^{-1} \Fh_{ij} U &=& F_{ij}\quad, \label{Ff1} \\
U^{-1}\Fh_{iz} U &=& D_i \l - [f,A_i]\quad, \\
U^{-1}\Dh_i \phih U &=& D_i \phi\quad, \\
U^{-1} \Dh_z \phih\,U &=& [\l\,,\phi] +[f,\phi]\quad,\label{Ff4} \\
U^{-1} \del_\mu \Ah^\mu\,U &=& \del_i A_i +[f,\lambda] \quad.  
\eea
Hence the gauge-fixed reduced action is given by
\bea \label{redact}
S_2 = \int \!\!\! &\Bigl\{& \!\!
(\del_i \phi)^2+ [f,\phi]^2  +(\del_i \l)^2 +[f,\l]^2 +
F_{tx}^2 + (\del_i A_i)^2 
+ [f, A_t]^2 +  [f, A_x]^2 
 \nn\\[4pt]
&&\!\! +\ 2 \del_i \phi [A_i,\phi] +2 \del_i \l[A_i,\l] 
-2[A_i,\l][f,A_i]
+2[f,\phi][\l,\phi] \\[4pt]
&&\!\! +\ [A_i\,,\phi]^2 +[A_i\,,\l]^2+ [\phi\,,\l]^2
\ \Bigr\} \quad. \nn
\eea
The first line represents the kinetic terms, while the second and third
lines are the interactions.
At this stage we further restrict our field configurations by 
imposing reality conditions on the reduced fields, namely
\be \label{real}
\phi^\dagger = \phi \quad,\qquad 
\lambda^\dagger = \lambda
\qquad\textrm{and}\qquad 
 A_i^\dagger = - A_i 
\quad.
\ee
This renders the action real and reduces the gauge group 
to $\textrm{U}_\star(1)$.

Summarizing, our linear background reduction~\eq{red-gen} takes the form
\be \label{ncred}
\psih = \ep^{z f}\,\psi\,\ep^{-z f} 
\qquad \textrm{with} \quad f = \a t + \b x
\qquad \textrm{for} \quad \psih = \phih, \im\Ah_\m \quad\text{real}\quad .
\ee
Note that the $z$-independence of the reduced fields $\psi$ implies 
the linear derivative constraint
\be \label{linear}
\del_z \psih = [f,\psih] = \im \th (\a \del_x - \b \del_t)\psih \quad,
\ee
where we have used 
\be
[t\, , \;] = \im \th \del_x \qquad\textrm{and}\qquad
[x\, , \;] = -\im \th \del_t \quad.
\ee
Likewise, the reality condition~\eq{real} lifts to 
\be \label{3dreal}
\psih^\dagger = \ep^{-z (f+f^\dagger)}\,\psih\,\ep^{z (f+f^\dagger)}
\ee
for the three-dimensional fields.
It is noteworthy that the two conditions \eq{linear} and~\eq{3dreal} 
are consistent. 

Now substituting the relation
\be
\del_z \psih = U [f,\psi] U^{-1}
\ee
in the action~\eq{redact},  its kinetic terms read
\be \label{KE-term}
-\phi \tilde{\D} \phi  
- A_i  \tilde{\D} A_i - \l \tilde{\D} \l \quad,
\ee
where the kinetic operator 
\be
\tilde{\D} := h^{ij} \del_i \del_j
\ee
contains the (inverse) metric 
\be \label{hij}
\begin{pmatrix} h^{tt} & h^{tx} \\[6pt] h^{xt} & h^{xx} \end{pmatrix} =
\begin{pmatrix} 1{-} \b^2 \theta^2 & \ \a \b\,\theta^2 \\[6pt]
                \a \b\,\theta^2 & 1{-} \a^2 \theta^2 \end{pmatrix}\quad.
\ee
Thus the reduction of the metric now reads
\be
\d_{\m\n} \to h_{ij} \quad.
\ee
Note that the signature of the metric depends on the values of $\a$ and $\b$. 
For the metric to be real, it is necessary that $\a$ and $\b$ be real or
purely imaginary.
In particular, the metric is Minkowskian if
\be \label{para}
(\a^2 +\b^2)\,\theta^2\ >\ 1 \quad.
\ee
Therefore a temporal direction can be generated in the lower dimension
if the parameters of our reduction (\ref{ncred}) obey (\ref{para})! 
It is clear 
that \eq{para} can never be satisfied in the commutative limit.
In fact, for $\theta\to0$ our reduction \eq{ncred} goes back to the
naive reduction~\eq{naive}.

We  note that unlike in ordinary Minkowskian gauge theory where the time
component of the gauge field has a kinetic term of the wrong sign, there is no
such problem in our case, see \eq{KE-term}. The difference is due to the 
appearance of gauge-fixing term $(\del_i A_i)^2$ in  
\eq{redact}. In ordinary gauge theory, 
this gauge-fixing term is not acceptable since it is not compatible with
the Lorentz symmetry SO$(1,1)$. In the noncommutative case, however,
the Lorentz symmetry breakdown renders this term admissible.

We remark that our reduction~\eq{ncred} resembles the one used in \cite{CL1}.
There the form of the reduction was fixed by requiring integrability, 
and it implied a change of signature from Lorentzian to Euclidean. 
This kind of signature change is typical for conventional dimensional 
reductions. Here we find that, by performing a more general dimensional 
reduction, one can do the opposite: changing of signature from Euclidean 
to Lorentzian. This is a novel phenomenon.

One may think that a constraint of the form~\eq{linear}, 
\be \label{linear1}
(\del_z + \im a\,\del_x + \im b\,\del_t )\,\psih(t,x,z) =0 
\qquad\textrm{with}\quad a,b \in\RR \quad,
\ee 
can be imposed in a commutative theory for achieving the same change of 
signature without invoking noncommutative geometry. However, 
it is easy to see that this is not the case: 
The complex nature of the constraint renders its resolution
in terms of real commuting fields impossible.
From the above analysis, we see that 
what noncommutative geometry effectively does for us
is to allow for a satisfactory implementation of the constraint~\eq{linear}. 
This brings about the desired change of signature
and would not be possible without noncommutative geometry at work.

The two key assumptions which enable this mechanism are, first, the reduction 
along some commutative direction in a noncommutative geometry and, second, 
the linear background reduction~\eq{ncred}. While the first ingredient is 
essential, the second one is open to generalization. The linear background
reduction~\eq{ncred} arose from the choice~\eq{UU}, which is the simplest one.
It is conceivable that more general forms of~$U$ in the twisted reduction
ansatz~\eq{red-gen} work as well. 
Another obvious generalization consists in adding further (commutative or
noncommutative) dimensions. Here, we have considered only the simplest case
of $D{=}2$, which we think is generic. The existence of commutative coordinates
in a Moyal-deformed geometry requires either a special choice of the
noncommutativity matrix~$(\theta^{\mu\nu})$ or an odd dimensionality of our 
space. In string theory, such a situation is realized on D-branes in the 
presence of a suitably chosen $B$-field background. 

Our lower-dimensional theory features a noncommutative time variable. 
In terms of the Moyal star product, the equation of motion contains an 
infinite number of time derivatives. The quantization of such theories is 
an unresolved problem (see for example
\cite{woodard,bahns,rimyee,sibold1,sibold2} for related discussions).
Yet we think that time-space noncommutative theories 
can make sense quantum mechanically.
A proper understanding of their quantum properties might in fact help us 
to better grasp the nature of spacetime in quantum gravity. 

Our reduction is parameterized by $\a$ and $\b$
which describe the background on which the reduction takes place. 
Just as for the case of an open string in a background NS-NS $B$-field, 
where the closed string metric and the background $B$-field are free and any
value of the noncommutativity parameters $\th^{\m\n}$ is allowed, 
the background parameters $\a$ and $\b$ can be dialed freely in our setup. 
Different choices give rise to lower-dimensional worlds with different 
``spacetime'' metrics, including the untwisted reduction as a special case. 
To decide which values for $\a$ and $\b$ are physically preferred,
an underlying ``microscopical'' theory is desired, which generates the
reduction via some compactification mechanism and lifts the flat
directions in our moduli space of reductions. This is beyond our reach at the 
moment. Clearly, it will be very interesting to obtain a dynamical/statistical
understanding of the origin of these parameters and the emergence of time. 

One of the most exciting prospects in the AdS/CFT proposal \cite{ads} 
is the possibility to understanding properties of spacetime in terms
of the dual gauge theory dynamics. It will be tantalizing to
understand how a change of spacetime signature manifests itself on the
gauge theory side in the context of AdS/CFT. It will also be very interesting
to investigate further the phenomenological implications of models
built on this kind of dimensional reduction. We leave these fascinating
issues for further investigation. 

\noindent
{\bf Acknowledgements\ }\\  
CSC would like to thank Pei-Ming Ho, Miao Li and Washington Taylor 
for helpful discussions and comments. 
OL thanks Alexander D.~Popov for remarks.
CSC acknowledges the support of EPSRC through an advanced fellowship.
The work of OL is partially supported by the Deutsche Forschungsgemeinschaft
(DFG).

\end{document}